# Supporting Information

## Porous Silicon Microcavities Coupling Resonant Raman Enhancement with Light-Driven Biomolecular Redox Control

*Irina S. Kriukova*[†,‡], *Almar Al-Assaad* [†], *Evelyn Granizo*[‡], *Aleksandr A. Knysh*[‡], *Daria Kalenichenko*[†], *Abdelilah Beljebbar*[†], *Alyona Sukhanova*[†], *and Igor Nabiev* [†,‡,*]

[†] Université de Reims Champagne-Ardenne, BioSpecT UR-7504, 51100 Reims, France

[‡] Nano-Photon Research Center, National Research Nuclear University MEPhI (Moscow Engineering Physics Institute), 115409 Moscow, Russia

* Corresponding author: Igor Nabiev, igor.nabiev@univ-reims.fr

## Supporting Note 1. A Physicochemical Model for the Photoinduced Reduction of Cytochrome C in a Porous Silicon Microcavity

### Theoretical model

The experimentally observed dependence of the fraction of reduced cytochrome *c* (Cyt*c*) on laser power exhibits pronounced saturation behavior. At first glance, this dependence could be approximated using the Michaelis–Menten equation [1] or a Langmuir-type function [2]:

(1)

where  is the degree of Cyt*c* reduction, *k* is the effective Cyt*c* reduction constant, and *P* is the laser irradiation power.

However, the direct applicability of the classical Langmuir formalism to the present system is not obvious because irradiation power cannot be interpreted as the concentration of an adsorbed or reacting species. In particular, laser power cannot simply replace the concentration term in the Langmuir equation by analogy. Therefore, rather than assuming a Langmuir-type dependence a priori, we derive the experimentally observed saturation behavior from the underlying photophysical and photochemical processes in the porous silicon (pSi)–Cyt*c* system. Specifically, the model accounts for photocarrier generation in pSi, electron–hole recombination, electron transfer to Cyt*c*, and competing oxidative pathways.

The primary objective is therefore not to derive a Langmuir-type equation per se, but to determine whether a physically justified kinetic description naturally gives rise to a saturating dependence of the reduced Cyt*c* fraction on laser power. The model can be represented by the following kinetic scheme, which captures the principal steps of the proposed photocatalytic mechanism.

***Photo-excitation***

,

(2)

where  is the incident energy, is a photoinduced electron,  is a photoinduced hole,  is the rate of generation of electron-hole pairs,  is the electron-hole pair generation constant, and  is the irradiation power.

***Recombination***

(3)

where  is the recombination rate and  is the recombination constant.

***<u>Capture of holes by the donor (D) or surface traps</u>***

The D can be either a solvent or an pSi surface state. This stage is crucial because holes can compete with the reduction and even reoxidize Cyt*c* ($Fe^{2+}$):

(4)

where  is the hole capture rate and  is the hole trapping constant.

***<u>Electron capture by surface traps</u>***

(5)

where  is the electron trapping rate and  is the electron trapping constant.

***Reduction of Cytc by photoelectrons***

(6)

where is the Cyt*c* reduction rate and is the effective constant of electron transfer from pSi to Cyt*c*.

***Reoxidation of reduced Cytc***

Reoxidation may be mediated by molecular oxygen, reactive oxygen species, photogenerated holes, surface states, or other electron acceptors. This process can be described as

(7)

where is the cytochrome oxidation rate, is the effective constant of Cyt*c* oxidation.

***Correction for irreversible losses***

(8)

where is the Cyt*c* loss rate and is the effective Cyt*c* loss constant.

Assuming that steady-state concentrations of photogenerated electrons and holes are established under continuous illumination [3], we obtain

, (9)

, (10)

From Eq. (10), it follows that

(11)

Substitution of this expression into Eq. (9) gives

(12)

For simplicity, we assume that the concentration of oxidized Cyt*c* available for reduction remains approximately constant over the timescale considered. Then, where .

Multiplying by the denominator and rearranging yields a quadratic equation for .

,

Solving this equation gives the physically meaningful positive root:

(13)

Thus, the steady-state photoelectron concentration is

(14)

.

This expression has the same mathematical structure as that encountered in Langmuir–Hinshelwood-type descriptions [4], although here it arises from the steady-state balance of photogenerated charge carriers rather than from an adsorption equilibrium. We then assume that

(15)

,

,

(16)

From Eqs. (15) and (16), we obtain

. (17)

For a first-order reaction, it follows from Eqs. (7), (8), (15), and (17) that

, (18)

Under continuous irradiation, following the initial transient, the system reaches a stationary photochemical state, as confirmed experimentally by the establishment of a stable Cyt*c* reduction level at each irradiation power. In this regime, the photogenerated-electron-driven reduction of Cyt*c* is dynamically balanced by the competing oxidative and loss processes, such that the reduction degree becomes time-independent on the experimental timescale:

Within the present kinetic model, this stationary condition corresponds to a balance between photogenerated-electron-driven reduction of Cytc and the competing oxidative and loss processes. Applying this condition to Eq. (18) yields:

, (19)

where .

To relate the theoretical expression to the experimentally measured quantity, a scaling factor, $\eta_{max}$, is introduced. In the present experiments, the degree of Cyt*c* reduction is evaluated from the $I_{750}/I_{1635}$ intensity ratio. The experimentally relevant function therefore takes the form:

(20)

where $\eta_0$ is the initial value of the Cyt*c* reduction parameter and $\eta_{max}$ is the scaling factor relating the theoretical reduction parameter $\eta(P)$ to the experimentally observed change in the $I_{750}/I_{1635}$ ratio.

**Model analysis**

Mathematically, the resulting expression, Eq. (20), has the form of a Langmuir-type saturation function and resembles the Michaelis–Menten equation [1]. In the present case, however, this functional form is obtained from physical considerations that differ fundamentally from those underlying classical Langmuir adsorption. Consequently, the experimentally observed saturation should not be interpreted merely as an empirical Langmuir-type approximation.

In the proposed model, the Langmuir-like functional form emerges naturally from the kinetic balance among photoelectron generation, electron–hole recombination, electron transfer to Cyt*c*, and competing oxidative pathways. The relevant effective quantity is the steady-state concentration of photogenerated electrons available for Cyt*c* reduction. Because this concentration is controlled by optical excitation and therefore depends on the incident laser irradiance, the final expression can be written as a function of laser power, the experimentally controlled variable. The resulting dependence thus retains the mathematical structure of a saturation function, while its parameters acquire physical meanings associated with photocarrier generation, recombination, trapping, and electron transfer in the pSi–Cyt*c* system.

Using the first-order Taylor approximation  for $|\alpha P| \ll 1$, Eq. (20) can be approximated as:

(21)

Furthermore, in the intermediate regime where $2|\beta| \ll |\alpha P| \ll 1$, the condition  allows Eq. (21) to be further expanded to first order in $2|\beta| / |\alpha P|$, yielding:

(22)

Importantly, Eq. (22) should not be interpreted as the high-power asymptotic limit of Eq. (20), because its derivation relies on the condition $|\alpha P| \ll 1$. The actual asymptotic behavior for high irradiation power ($P \to \infty$) must therefore be obtained directly from Eq. (20). Since at large $P$. Eq. (20) can be written as:

(23)

Accordingly, the approach to saturation follows a $P^{-1/2}$ dependence rather than $P^{-1}$ dependence that would result from extrapolating the intermediate-regime approximation. Thus, at sufficiently high irradiation power, the degree of Cytc reduction approaches the limiting value , while the residual deviation from saturation decreases as . Physically, this behavior reflects progressively diminishing changes in the degree of Cytc reduction with further increases in irradiation power.

To quantitatively characterize the transition toward saturation, we introduce the characteristic half-saturation power ($P_{1/2}$) defined as the irradiation power at which the photoinduced increase in the degree of Cyt*c* reduction reaches one-half of its maximum value:

(24)

If , then

(25)

Eq. (20) can also be linearized:

(26)

. (27)

Thus, the model predicts a linear relationship in the transformed coordinates and .

In a more general physical case, ; therefore,

(28)

**Experimental data analysis: comparison of the theoretical model with the Michaelis–Menten equation**

To demonstrate the robustness of the proposed model, an experimental dataset exhibiting greater scatter was analyzed in detail in Supporting Note 1. Despite the increased experimental variability, the fitting procedure yielded physically meaningful parameter values, supporting the applicability of the model under less favorable experimental conditions. The theoretical model yields the following approximation function for :

(29)

The Michaelis–Menten approximation function for  is

. (30)

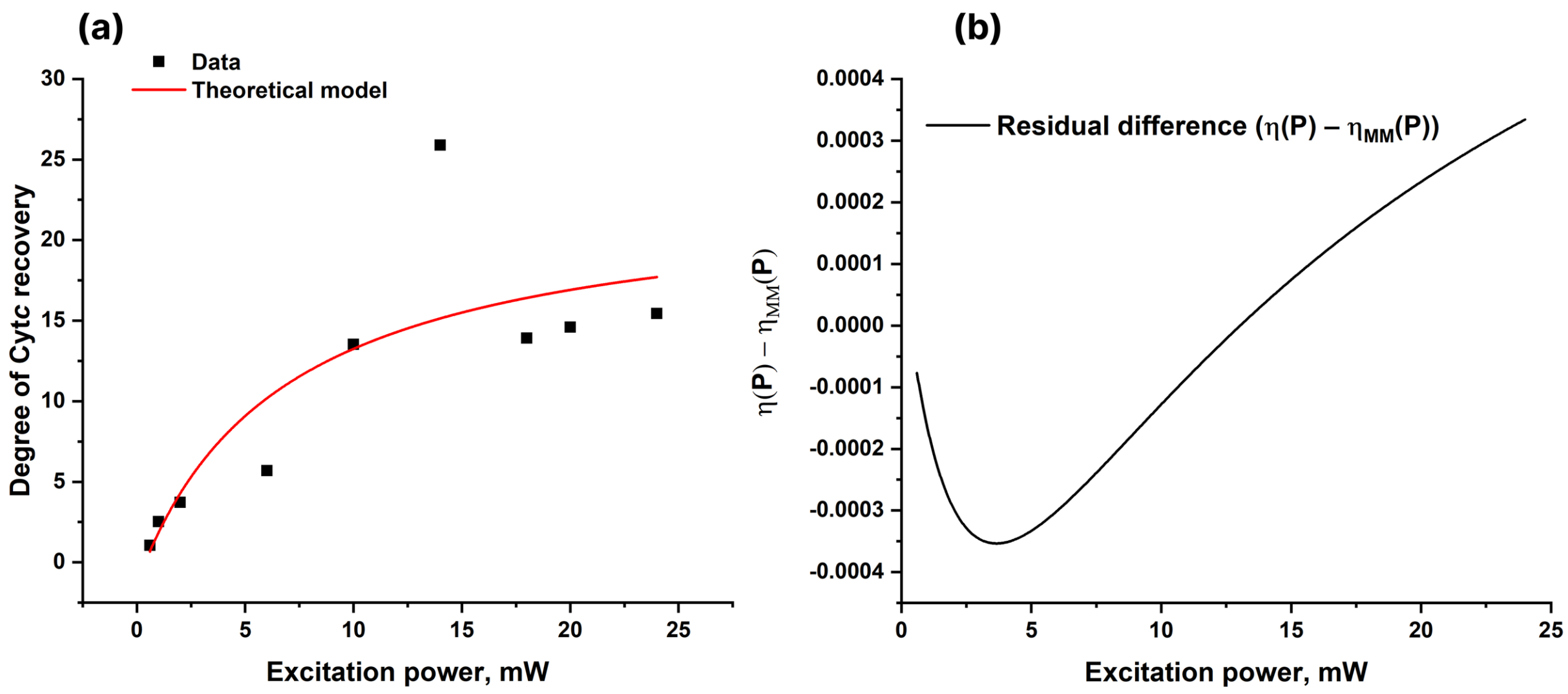

**Figure S1.** Comparison of the theoretical model and the Michaelis–Menten description of cytochrome c (Cyt*c*) reduction in a porous silicon microcavity. (a) Experimental data and approximation by the proposed theoretical model. (b) The residual difference between the predictions of the proposed model and the Michaelis–Menten model.

**Figure S1** shows excellent agreement between the two models over the investigated range of irradiation powers. This close correspondence demonstrates that the Cyt*c* reduction process can also be represented mathematically using a Michaelis–Menten-type formalism, provided that its parameters are interpreted in terms of the relevant photophysical quantities rather than conventional enzyme-kinetic variables. The difference between the predictions of the two models remains below $10^{-4}$, confirming the near-equivalence of their functional forms under the experimental conditions considered.

To ensure physical consistency, we assume  in both models. Then, our theoretical model yields the following approximation function for :

(31)

and the Michaelis–Menten approximation function for is

(32)

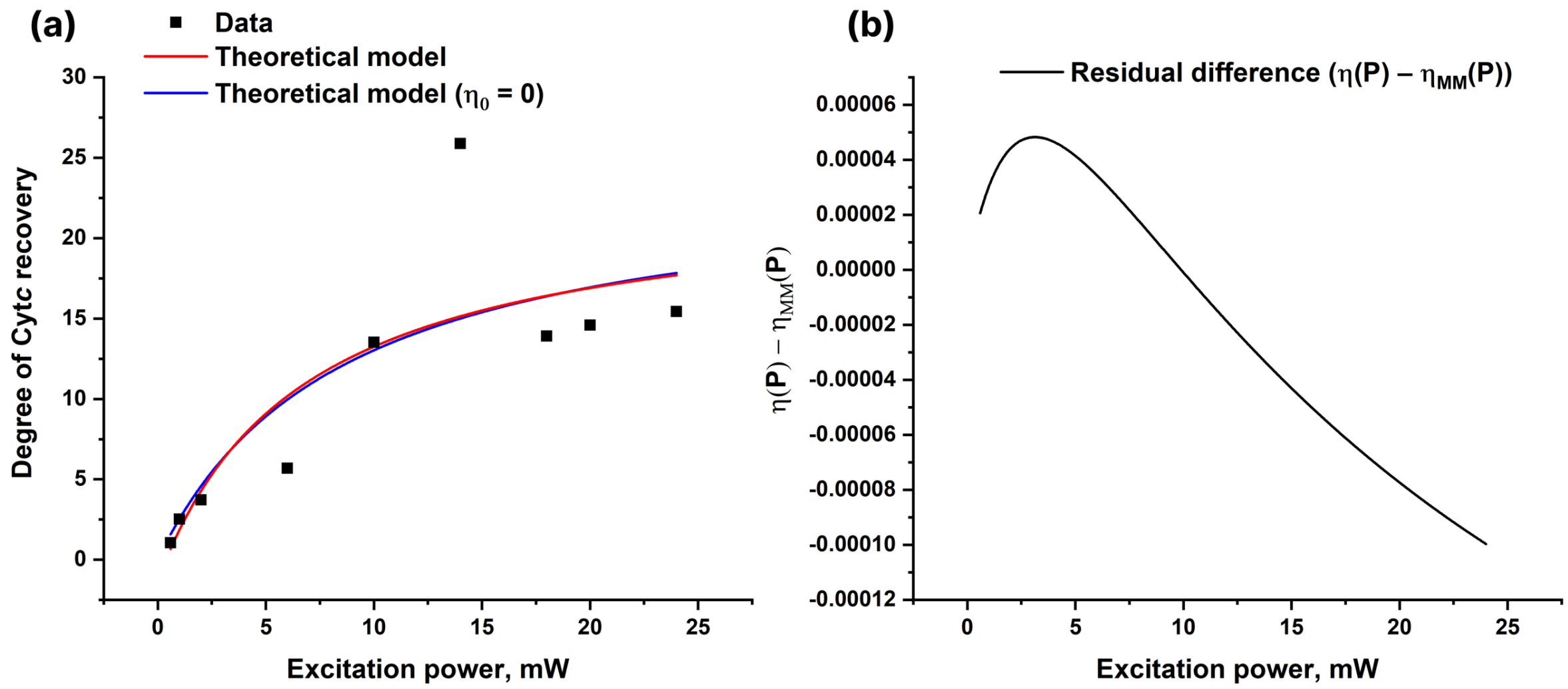

**Figure S2.** Comparison of the theoretical model and the Michaelis–Menten description of cytochrome c (Cyt*c*) reduction in a porous silicon microcavity for . (a) Experimental data and approximation by the proposed theoretical model with (blue). (b) Residual difference between the predictions of the proposed theoretical model and the Michaelis–Menten model under the same constraint.

**Figure S2** again demonstrates excellent agreement between the two models over the entire range of irradiation powers investigated. Under this constraint, the discrepancy between their predictions is even smaller than that shown in **Figure S1**, remaining below $10^{-5}$. This close agreement indicates that, within the experimentally accessible power range, the two functional forms are essentially indistinguishable. Importantly, however, the proposed model provides a physicochemical interpretation of saturation behavior in terms of photocarrier generation, recombination, trapping, electron transfer, and competing oxidation processes.

**Experimental data analysis: calculation of the characteristic half-saturation power ($P_{1/2}$)**

Note that for

Approximation of experimental data by Eq. (29) yields $\alpha = 2.606609 \cdot 10^{-5}$ and $\beta = 8.737576 \cdot 10^{-5}$. Therefore,

If = 0, the same approximation yields $\alpha = 1.972351 \cdot 10^{-5}$ and $\beta = 8.472002 \cdot 10^{-5}$; hence,

For the present system, can therefore be estimated from these two limiting fits as an average of these two values: .

## Experimental data analysis: linearization of the theoretical model

**Figure S3** shows the experimental data transformed according to Eqs. (27) and (28), using the fitting parameters obtained from Eqs. (29) and (31).

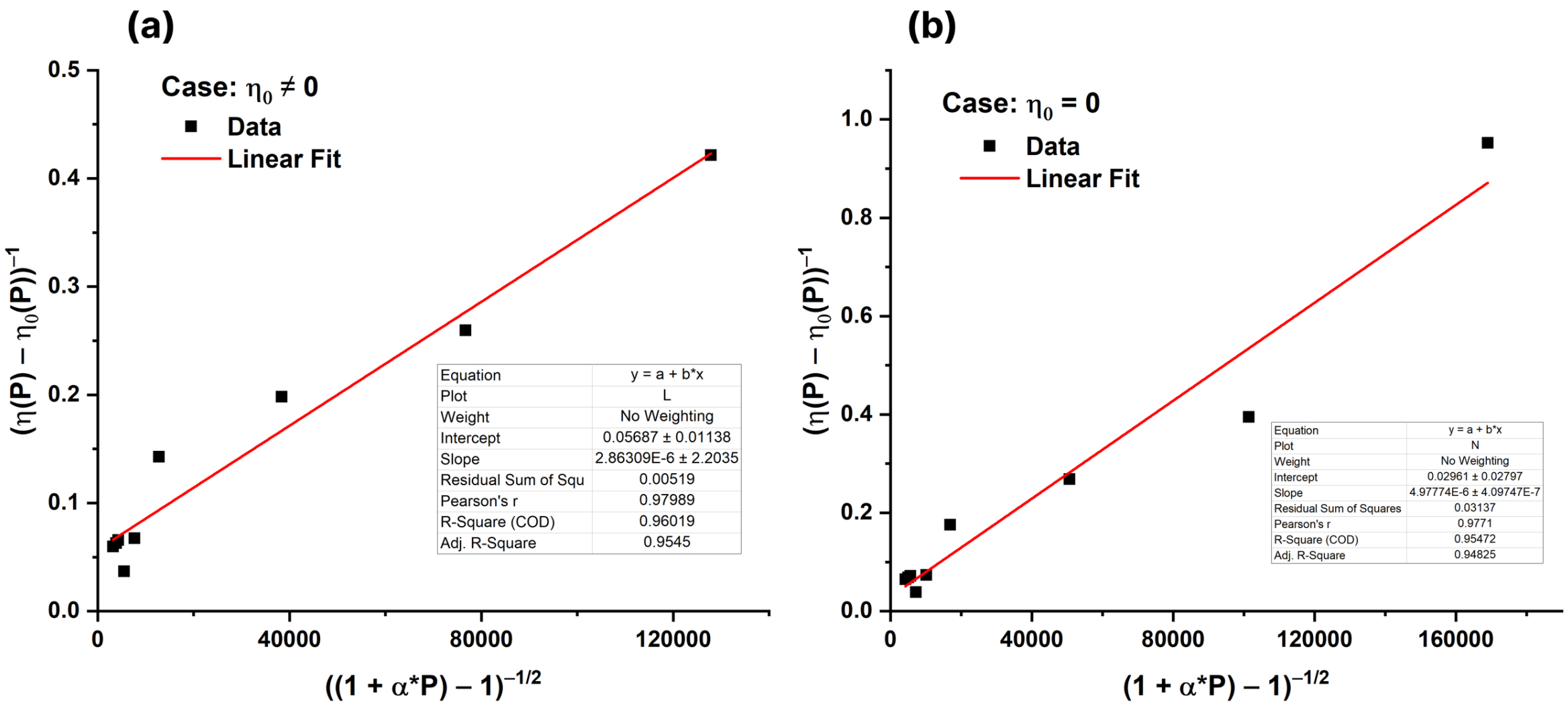

**Figure S3.** The experimental data transformed according to Eqs. (27) and (28), using the fitting parameters obtained from Eqs. (29) and (31). (a)  and (b) .

The linear behavior of the transformed experimental data provides additional support for the proposed model. In particular, the agreement between the predicted and experimentally observed linearized dependences demonstrates the internal consistency of the derived analytical expression and its ability to reproduce the observed power-dependent saturation behavior. Together with the direct nonlinear fits shown in **Figures S1** and **S2**, this analysis supports the proposed physicochemical interpretation of photoinduced Cyt*c* reduction in the pSi microcavity.

Overall, the proposed physicochemical model provides a mechanistic description of the nonlinear, saturating dependence of Cyt*c* reduction on irradiation power. Within this framework, the observed saturation emerges naturally from the steady-state balance among photocarrier generation in pSi, electron–hole recombination and trapping, interfacial electron transfer to Cyt*c*, and competing oxidative pathways. Importantly, although the resulting analytical expression formally resembles Langmuir- or Michaelis–Menten-type saturation kinetics, its physical origin is fundamentally different: irradiation power governs the steady-state photocarrier population rather than representing the concentration of an adsorbed or reacting species. The model thus establishes a direct link between the experimentally observed power dependence and the underlying photophysical and interfacial charge-transfer processes. The agreement between the model and the experimental data, supported by the linearized representation and the predicted approach to a well-defined saturation limit, is consistent with a photocarrier-mediated mechanism underlying the light-driven $Fe^{3+} \rightarrow Fe^{2+}$ conversion of Cyt*c* within the pSi microcavity.

## Supporting Note 2. Numerical Simulations of the Optical Response and Field Confinement in a Porous Silicon Microcavity.

The electromagnetic response of the porous silicon microcavity was simulated using the Electromagnetic Waves, Frequency Domain (EMW) interface implemented in COMSOL Multiphysics 5.5. The simulated structure consisted of two distributed Bragg reflectors (DBRs) separated by a defect layer forming the optical cavity. The upper and lower DBRs comprised 6 and 10 alternating pairs, respectively, of low- and high-porosity layers (A and B, respectively), while the cavity was formed by a low-porosity layer with an optical thickness corresponding to 2A (**Figure S4**).

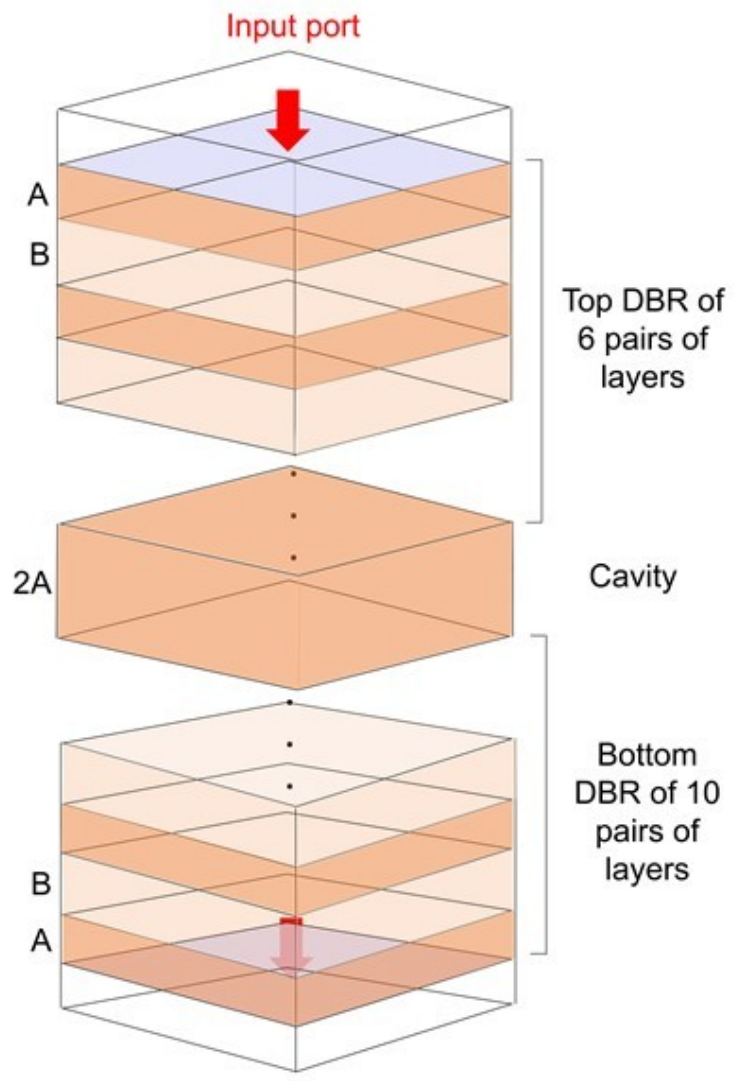

**Figure S4.** Schematic representation of the simulated porous silicon microcavity.

The simulations were performed using the finite element method (FEM). The refractive indices employed in the model were calculated for porous silicon layers fabricated at current densities of 5 and 38 mA/cm², corresponding to etching currents of approximately 3.5 and 26.8 mA, respectively, under the experimental electrode geometry [5].

Before Raman measurements of cytochrome c in porous silicon microcavities, the structures were thermally oxidized and subsequently incubated in an aqueous solution of cytochrome c. Both oxidation and infiltration of the porous structure by the aqueous medium modify the effective refractive indices of the constituent layers and, consequently, the optical response of the entire microcavity. To account for these changes, three states of the microcavity were simulated: (1) non-oxidized, (2) thermally oxidized, and (3) thermally oxidized and immersed in water.

The physical thicknesses of the layers were kept identical in all three models because they were determined by the electrochemical etching conditions. The low-porosity layers fabricated at a current density of 5 mA/ cm² (corresponding to an etching current of approximately 3.5 mA) and the high-porosity layers fabricated at 38 mA/cm² (approximately 26.8 mA) were assigned thicknesses of 63 and 97 nm, respectively. A port boundary condition was used to excite the structure with an incident electromagnetic wave. The input power at the excitation port was set to 1 W, and the electric-field mode amplitude was set to $E_x = 1$ V/m.

The three models differed in the refractive indices assigned to the porous layers to represent the successive stages of microcavity preparation. The first model represented the structure immediately after electrochemical etching, i.e., the non-oxidized porous silicon microcavity. The refractive indices of the low- and high-porosity layers were $2.345 - 0.01i$ and $1.516 - 0.002i$, respectively. The second model represented the thermally oxidized microcavity, for which the corresponding refractive indices were $1.913 - 0.006i$ and $1.392 - 0.002i$. In the third model, representing the oxidized porous silicon microcavity immersed in water, the refractive indices of the low- and high-porosity layers were $2.027 - 0.005i$ and $1.661 - 0.002i$, respectively.

The simulated transmission spectra exhibited cavity resonances at 588, 505, and 570 nm for the non-oxidized, thermally oxidized, and oxidized water-immersed microcavities, respectively (**Figure S5**). The pronounced spectral shifts between the three states reflect changes in the effective optical thickness of the multilayer structure resulting from oxidation and infiltration of the porous medium by water.

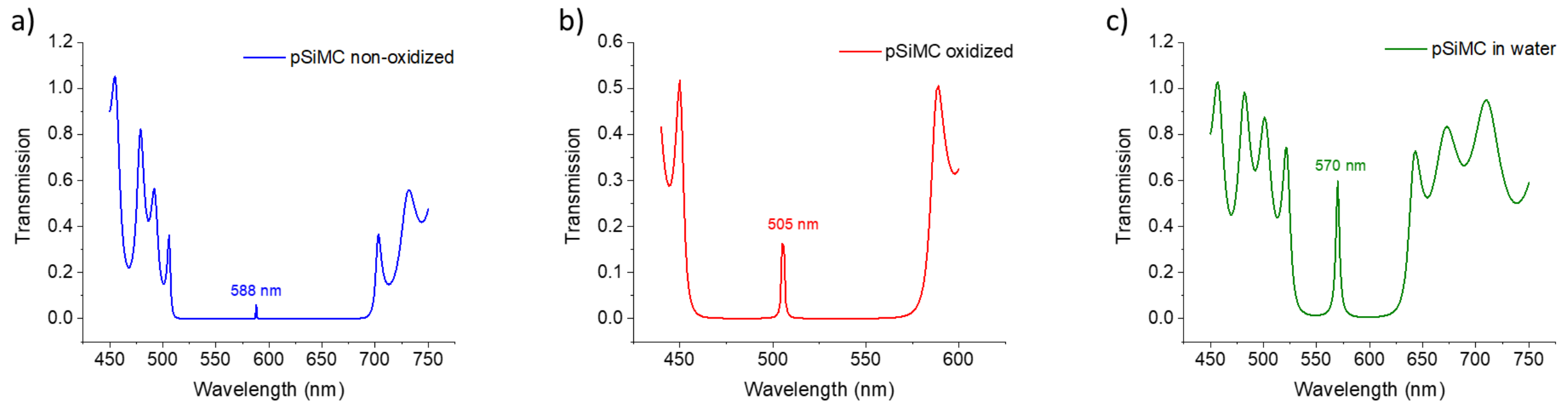

**Figure S5.** Simulated transmission spectra of (a) the non-oxidized porous silicon microcavity, (b) the thermally oxidized microcavity, and (c) the oxidized microcavity immersed in water. The corresponding cavity-resonance wavelengths are 588, 505, and 570 nm, respectively.

The spatial electric-field distribution, representing the optical mode profile within the multilayer structure, was evaluated at the resonance wavelength of each microcavity using the built-in COMSOL variable emw.normE. In all three cases, the simulations revealed qualitatively similar field-distribution patterns, characterized by pronounced localization of the electromagnetic field within the cavity region. As a representative example, **Figure S6** shows the electric-field distribution calculated for the oxidized porous silicon microcavity immersed in water.

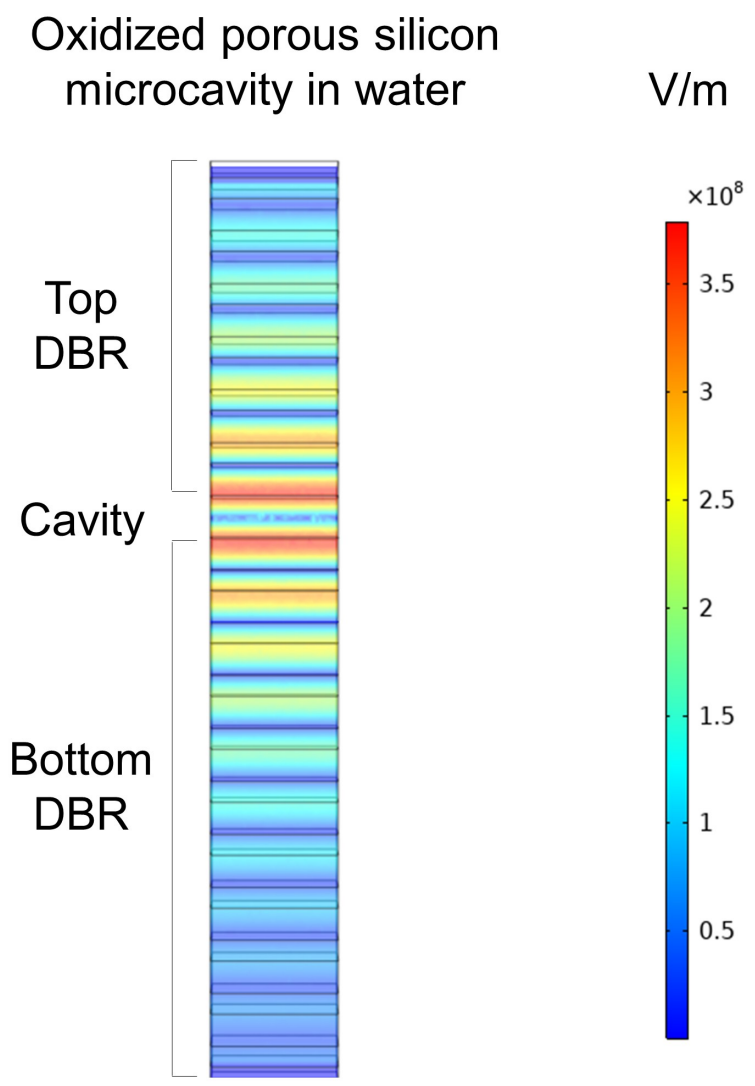

**Figure S6.** Simulated spatial distribution of the electric-field magnitude in the oxidized porous silicon microcavity immersed in water at the cavity-resonance wavelength, λ = 570 nm.

The calculated field distributions were further analyzed at selected positions across the cavity region, as shown in **Figure S7**. The inset schematically indicates the cross-section of the microcavity used for this analysis: the cavity region is shown in purple, the sampling positions are indicated by dots, and the coordinate x = 0 corresponds to the center of the cavity.

For all three simulated microcavity states, the electric-field distribution within the cavity was spatially nonuniform, with local maxima occurring near the interfaces between the defect layer and the adjacent DBRs. This behavior is characteristic of the resonant optical mode established within the multilayer cavity. Differences among the three calculated field profiles arise primarily from changes in the refractive-index contrast between the low- and high-porosity layers.

Among the three structures considered, the non-oxidized microcavity exhibited the largest refractive-index contrast between the constituent porous layers, whereas the oxidized microcavity immersed in water exhibited the smallest contrast. These differences modify the reflectivity of the DBRs and therefore affect both the spectral characteristics of the cavity resonance and the spatial confinement of the electromagnetic field. Consistent with this interpretation, the non-oxidized microcavity exhibited a narrower calculated resonance, indicative of stronger optical confinement and greater spectral selectivity. In contrast, the oxidized water-immersed microcavity exhibited a broader and more pronounced transmission resonance, consistent with reduced DBR reflectivity and weaker confinement of the cavity mode.

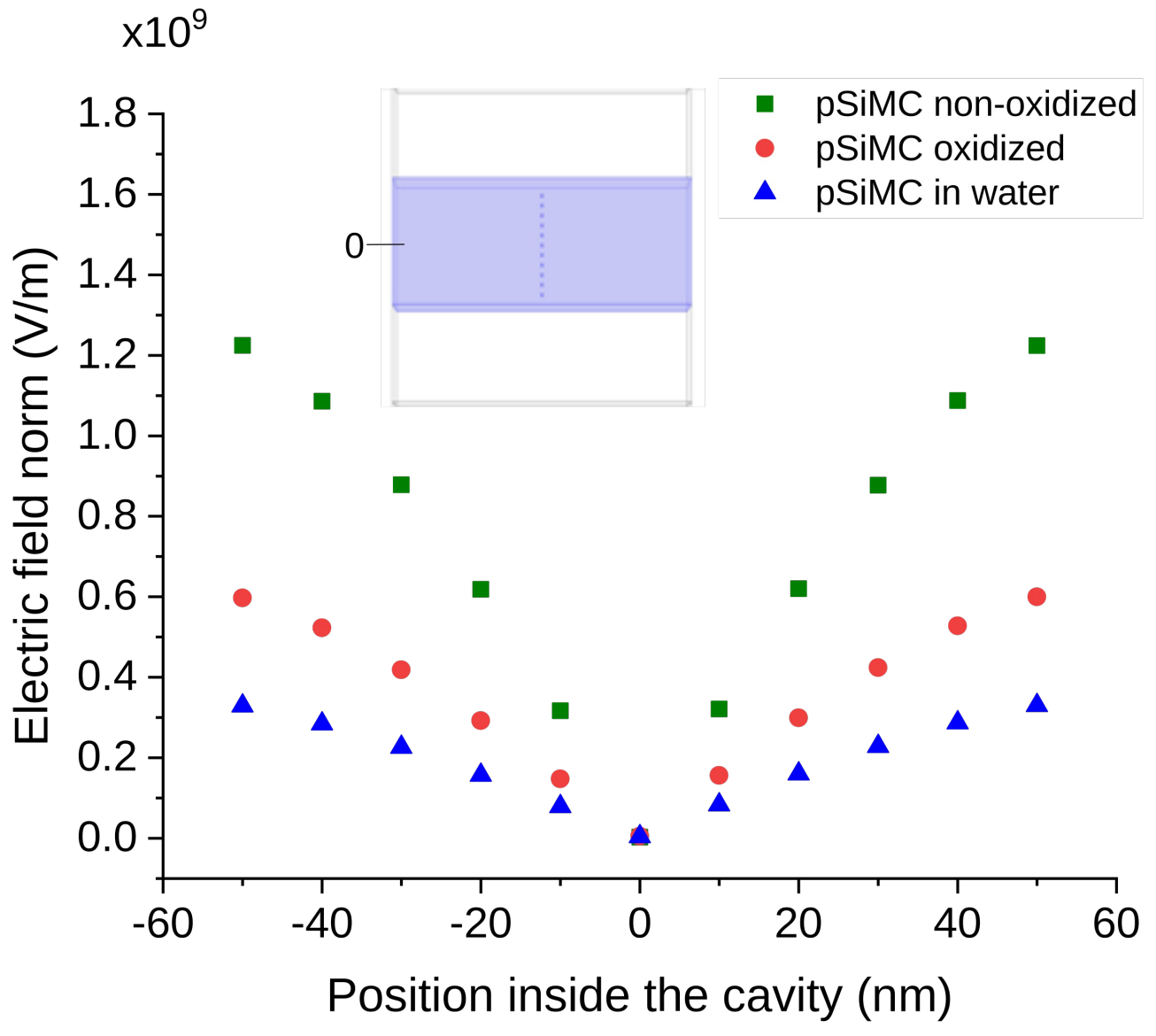

**Figure S7.** Comparison of the simulated electric-field distributions within the non-oxidized porous silicon microcavity (pSiMC non-oxidized), the thermally oxidized microcavity (pSiMC oxidized), and the oxidized microcavity immersed in water (pSiMC in water).
The coordinate x = 0 corresponds to the center of the cavity. The inset indicates the positions at which the electric-field magnitude was evaluated across the cavity region.

The electromagnetic simulations demonstrate that the optical response of the pSi microcavity is strongly affected by oxidation and infiltration of the porous network by the aqueous medium, which modify the effective refractive indices of the constituent layers and shift the cavity resonance. Importantly, for the experimentally relevant oxidized, water-infiltrated structure, the calculated resonance occurs at approximately 570 nm, in agreement with the spectral region targeted for enhancement of the Raman-scattered light. The calculated field distributions further demonstrate spatial localization and nonuniform enhancement of the electromagnetic field within the defect region of the microcavity. These results provide an electromagnetic basis for interpreting the experimentally observed enhancement of Cyt*c* Raman scattering as a consequence of resonant photon confinement and increased light–matter interaction within the pSi microcavity.

## Supplementary References